 \documentclass[final,1p,times]{elsarticle}
\usepackage{booktabs}
\usepackage{nomencl} % makeindex CEP-D-11-00194.nlo -s nomencl.ist -o CEP-D-11-00194.nls
\usepackage{acronym} % list of acronyms in the text
\makenomenclature

\journal{Computers \& Fluids}

\begin{document}

\begin{frontmatter}



\title{Lattice Boltzmann method for shape optimization of fluid distributor}

\author[label1]{Limin Wang \corref{cor1}}
\ead{lmwang@home.ipe.ac.cn} \cortext[cor1]{Corresponding author.
Tel.: +86 10 8254 4942; fax: +86 10 6255 8065.}
\author[label2]{Yilin Fan}
\author[label2]{Lingai Luo \corref{cor1}}\ead{Lingai.Luo@univ-nantes.fr}
\address[label1]{State Key Laboratory of Multiphase Complex Systems, Institute of Process Engineering, Chinese Academy of Sciences, Beijing 100190, China}
\address[label2]{Laboratoire de Thermocin\'{e}tique de Nantes, UMR CNRS 6607, Polytech' Nantes - Universit\'{e} de Nantes, La Chantrerie, Rue Christian Pauc, BP 50609, 44306 Nantes Cedex 03, France}
\begin{abstract}
This paper presents the shape optimization of a flat-type arborescent fluid distributor for the purpose of process intensification. A shape optimization algorithm based on the lattice Boltzmann method (LBM) is proposed with the objective of decreasing the flow resistance of such distributor at the constraint of constant fluid volume. Prototypes of the initial distributor as well as the optimized one are designed. Fluid distribution and hydraulic characteristics of these distributors are investigated numerically. Results show that the pressure drop of the optimized distributor is between 15.9\% and 25.1\% lower than that of the initial reference while keeping a uniform flow distribution, demonstrating the process intensification in fluid distributor, and suggesting the interests of the proposed optimization algorithm in engineering optimal design.
\end{abstract}


\end{frontmatter}

% \linenumbers
%% \iffalse
%% \fi

\section{Introduction}
\label{}
Delivering and distributing flows of one or different fluids precisely onto a given surface or into a given volume is an important issue in many unit operations of process engineering. For example, homogeneous flow distribution is generally required for multi-channel heat-exchangers [1-3]; for solar collectors [4-6]; for trickle bed, catalytic packed bed or multi-channel reactors [7-12]; for catalytic monoliths [13,14] or for bubble column [15,16].

It has been well recognized that the flow distribution uniformity is generally associated with proper design of entrance and exit manifolds, or the configuration of fluid distributors/collectors [17,18]. Many researches have focused on the design and optimization of fluid distributors/collectors, in particular, arborescent architectures have been given much attention; whether from one point to a line [9,12,19,20] or to the periphery of a disc [21-23]; to a surface [24-27] or to a volume [28,29]. The basic architectures of multi-scale channel networks build on analogies with living organisms (e.g. the lung or the vascular system). Design and optimization procedures have been proposed by Tondeur and his coworkers [26,27,29,30] to minimize the pressure drop (viscous dissipation) under constraints of uniform irrigation and constant void volume based on the principle of ``equipartition properties", which is in close connection to the constructal theory developed by Bejan [31-33]. Analytical, numerical and experimental investigations carried out by Luo and her coworkers [29,34,35] imply that arborescent structures that may guarantee the flow distribution uniformity between the channels will induce higher pressure drops, mainly because of the so-called ``minor losses" due to numerous singularities of the complex multi-scale structure (bifurcation, elbow, etc.). These losses are actually not so minor. As a result, for the design of an arborescent fluid distributor, can the analytical optimization be further improved, particularly for the profiling of the junctions in complex structures?

In the last decades, numerical techniques have been developed to solve the problem of flow shape optimization. Earlier work has studied various aspects of shape/topology optimization of fluid flow [36-38]. Noteworthy is the work of Errera and Bejan [39] which proved that the dendritic patterns formed by low-resistance channels in a river drainage basin can be deduced from the constrained minimization of global resistance in area-to-point flow. Moos and his coworkers [40,41] also proposed a procedure for topology optimization of fluid flow based on the principle that the fluid flow always searches for the best way under given constraints in a predefined space by itself.

So emerges the idea that classic optimization approaches could benefit of numerical computing to overcome the analytical difficulty of expressing objective function and constraints in complex geometries. In that context, we have developed a lattice Boltzmann method (LBM) shape optimization algorithm for minimizing the resistance of flow structures at the constraint of fixed void volume for fluid flow [42]. Numerical examples of a right angle elbow and a T-junction show that this algorithm can optimize the flow shape with significantly reduced flow resistance. In the present study, this algorithm is applied to optimize the structure of a flat-type arborescent distributor investigated in reference [27]. Numerical experiments are also carried out in parallel for comparing flow distribution and hydrodynamic characteristics of the distributors. Results obtained are useful for further validation of our optimization algorithm, showing its promising application in processes dealing with fluid.

\section{Numerical algorithm}
\label{}
LBM, originally proposed by McNamara and Zantti (1988) [43] as a smoothed alternative to lattice gas automata (LGA), is an efficient second-order Navier-Stokes solver capable of solving various systems for hydrodynamics owing to its algorithmic simplicity, explicit solution of particle distribution functions, natural parallelism and easy boundary treatment [44]. The objective of this algorithm is to optimize the shape of fluid flow by minimizing the flow resistance (thus the pressure drop) under the constraint of constant void volume for fluid flow. To do that, a 2D simulation domain including both fluid phase and solid phase is uniformly divided into elemental square cells. LBM is utilized as an underlying Navier-Stokes solver to calculate the flow flied. Compared to the traditional CFD methods based on discretizations of macroscopic continuum equations, the kinetic nature and local dynamics of LBM make it more adaptive to complex boundaries and parallel computing [44]. In addition, the theoretical basis (``cell" expression) of the LBM corresponds very well to the downstream cells' position exchanging procedure. That is the reason why we choose the LBM as the ``pretreatment".

To simplify the numerical algorithm, following assumptions are made:
\begin{enumerate}
\item  [-] Steady flow pattern; No-slip condition at the wall.
\item  [-] 2-D simulation, i.e. infinite channel depth; only consider the friction between the fluid and the solid walls at both sides; negligible gravity effect;
\item  [-] Isothermal operating condition; isotropic and homogeneous physical properties of solid materials; constant physical properties of working fluid.
\end{enumerate}

\subsection{Standard lattice Boltzmann formulations for fluid flows}
In standard lattice Boltzmann method, the macroscopic equations of traditional CFD (the Navier-Stokes equations) are not solved directly, but rather the Boltzmann equation is solved on a discrete lattice. For the solution of incompressible fluid flow, we use the classical D2Q9 scheme of LBM, with three speeds (0, 1 and $\sqrt{2}$) and nine velocities (${\bf{c}}_0 \sim {\bf{c}}_8$) on a two-dimensional square lattice (Fig. 1), the evolution of the distribution function $f_i \left( {{\bf{x}},t} \right)$ obeys the lattice Boltzman equation (LBE) [44]:
\begin{equation}\label{eq1}
f_i \left( {{\bf{x}} + {\bf{c}}_i \Delta t,{\rm{ }}t + \Delta t} \right) - f_i \left( {{\bf{x}},{\rm{ }}t} \right) =  - \frac{1}{\tau }\left( {f_i  - f_i^{\rm{eq}} } \right)
{\rm{,}}{\kern 1pt} {\kern 1pt} {\kern 1pt} {\kern 1pt} {\kern 1pt} {\kern 1pt} {\kern 1pt} {\kern 1pt} {\kern 1pt} {\kern 1pt} {\kern 1pt} {\kern 1pt} {\kern 1pt} {\kern 1pt} {\kern 1pt} {\kern 1pt} {\kern 1pt} {\kern 1pt} {\kern 1pt} {\kern 1pt} i{\rm{ = }}0{\rm{, }}1{\rm{, }}2{\kern 1pt} {\kern 1pt} {\kern 1pt} {\kern 1pt}  \cdots {\kern 1pt} {\kern 1pt} {\kern 1pt} {\kern 1pt} 8
\end{equation}
Here ${{\bf{c}}_i}{\rm{ = }}\left\{ {\left. {\cos \left[ {\left( {i - {\rm{1}}} \right)\pi {\rm{/}}2} \right], \sin \left[ {\left( {i - {\rm{1}}} \right)\pi {\rm{/}}2} \right]} \right\}} \right.$  for $i{\rm{=1}}\sim 4$, ${{\bf{c}}_i}{\rm{ = }}\sqrt 2 \left\{ {\left. {\cos \left[ {\left( {i - 5} \right)\pi {\rm{/}}2} \right], \sin \left[ {\left( {i - 5} \right)\pi {\rm{/}}2} \right]} \right\}} \right.$  for $i{\rm{= 5}}\sim 8$ and ${{\bf{c}}_0}{\rm{ = }}0$, and $\tau  = 3\nu  + \frac{{\Delta t}}{2}$ is the relaxation time related to kinematic viscosity $\nu$ and the discrete time step $\Delta t$, and $f_i^{{\rm{eq}}}$ is the equilibrium distribution function defined as [45],
\begin{equation}
  f_i^{{\rm{eq}}}\left( {\rho ,{\bf{u}}} \right) = {\omega _i}\rho \left( {1 + \frac{{{{\bf{c}}_i} \cdot {\bf{u}}}}{{c_{\rm{s}}^{\rm{2}}}} + \frac{1}{2}{{\frac{{\left( {{{\bf{c}}_i} \cdot {\bf{u}}} \right)}}{{c_{\rm{s}}^{\rm{4}}}}}^2} - \frac{1}{2}\frac{{{\bf{u}} \cdot {\bf{u}}}}{{c_{\rm{s}}^{\rm{2}}}}} \right){\rm{,}}{\kern 1pt} {\kern 1pt} {\kern 1pt} {\kern 1pt} {\kern 1pt} {\kern 1pt} {\kern 1pt} {\kern 1pt} {\kern 1pt} {\kern 1pt} {\kern 1pt} {\kern 1pt} {\kern 1pt} {\kern 1pt} {\kern 1pt} {\kern 1pt} {\kern 1pt} {\kern 1pt} {\kern 1pt} {\kern 1pt} {\kern 1pt} i{\rm{ = }}0{\rm{, }}1{\rm{, }}2{\kern 1pt} {\kern 1pt} {\kern 1pt} {\kern 1pt}  \cdots {\kern 1pt} {\kern 1pt} {\kern 1pt} {\kern 1pt} 8
\end{equation}
where the weights are given by ${\omega _0}{\rm{ = 4/}9}$, ${\omega _i}{\rm{ = }}1{\rm{/}}9$ for $i{\rm{=1}}\sim 4$, ${\omega _i}{\rm{ = }}1{\rm{/}}36$ for $i{\rm{= 5}}\sim 8$,  and $c_{\rm{s}}$ is the speed of sound and equals to $\sqrt{3}\rm{/}3$  in lattice unit, $\rho$ and \textbf{u} are density and velocity, respectively.

\begin{figure}
  \centering
  \includegraphics[width=8.3cm]{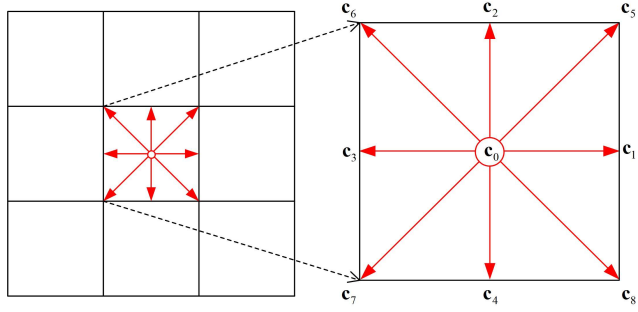}
  \caption{A two-dimensional (2D), 9-velocity D2Q9 lattice model}\label{fig1}
\end{figure}

The macroscopic variables such as fluid density and momentum are related to the distribution function and obtained as:
\begin{equation}\label{eq2}
\rho \left( {{\bf{x}},{\rm{ }}t} \right) = \sum\limits_i {f_i \left( {{\bf{x}},{\rm{ }}t} \right)} ,{\kern 12pt} \rho {\bf{u}}\left( {{\bf{x}},{\rm{ }}t} \right) = \sum\limits_i {{\bf{c}}_i f_i \left( {{\bf{x}},{\rm{ }}t} \right)}
\end{equation}

Through the Chapman-Enskog procedure, the incompressible Navier-Stokes equations can be obtained from the LBE in the limit of small Mach number [46-48], as the following:
\begin{equation}\label{eq3}
\begin{array}{l}
 \frac{{\partial \rho }}{{\partial t}} + \nabla  \cdot \left( {\rho {\bf{u}}} \right) = 0 \\
 \\
 \frac{{\partial \left( {\rho {\bf{u}}} \right)}}{{\partial t}} + \nabla  \cdot \left( {\rho {\bf{uu}}} \right) =  - \nabla \left( {\rho c_{\rm{s}}^2 } \right) + \nabla  \cdot \left( {\rho \nu \nabla {\bf{u}}} \right) + {\bf{F}} \\
 \end{array}
\end{equation}
where the pressure is given by:
\begin{equation}\label{eq4}
p = \rho c_{\rm{s}}^2
\end{equation}

\subsection{Mutual replacements between fluid and solid cells}
We pay close attention to the dynamic interaction at fluid-solid interface and propose a heuristic optimality criterion, i.e. the viscous stress for solid cells:
\begin{equation}\label{eq5}
\tau _{yx}  = \mu \frac{u}{{0.5\Delta y}},{\kern 12pt} \tau _{xy}  = \mu \frac{\upsilon }{{0.5\Delta x}}
\end{equation}
and dynamic pressure for fluid cells:
\begin{equation}\label{eq6}
q = \frac{1}{2}\rho \|\bf{u}\|^2
\end{equation}
where $\Delta x$ and $\Delta y$ are the lattice spacing ($\Delta x = \Delta y$ in the LBM), $u$  and $\upsilon$ are velocity in x and y direction, respectively.

In our optimized algorithm, solid cells at fluid-solid interface suffering from larger viscous stress from its neighboring fluid cells will vanish and be replaced by fluid cells. Likewise, fluid cells at the fluid-solid interface having lower dynamic pressure are replaced by solid cells, in order to eliminate the ``dead zones" in the fluid domain so as to effectively make use of a fixed void volume for fluid flow. Note that the equal number of fluid cells and solid cells is targeted to balance the void volume occupied by fluid. The mutual replacement of fluid and solid cells will create a new flow shape that differs from the initial one. This cells' position exchange process in our algorithm can be considered as a ``generalized Cellular Automaton" and used the term ``CA" for short. Then the flow field of this new shape will again be calculated by LBM for the recurrence by a CA procedure. Step by step, the shape of fluid flow evolves towards the final shape with homogenized dynamic pressure at fluid-solid interface and reduced total flow resistance.

From the view point of fluvial dynamics, our algorithm can be thought of as a mimicry of natural behavior in river channels where the surface is eroded at the points of maximum shear stress and the sand is deposited at the points of minimum dynamic pressure [42].

\subsection{Numerical implementations}
The optimization procedure is described in detail using the flow chart shown in Fig. 2.
\begin{enumerate}
  \item [(1)] Input the initial data such as the size and initial shape of the simulation domain (solid phase, fluid phase), the specified boundary conditions (fluid nature, velocity profile, pressure, etc.).
  \item [(2)] An exact flow field is calculated by LBM.
  \item [(3)] At the fluid-solid interface, a number of fluid cells having the lowest dynamic pressure and the same number of solid cells suffering the largest viscous stress will be identified, and their positions will be exchanged, thus creating a new shape.
  \item [(4)] Reinitialize the basic properties of fluid flow and boundary for the new shape. Recalculate the exact flow field by LBM.
  \item [(5)] Check the stable tolerance of the algorithm. If the tolerance is satisfied, then the heuristic procedure is terminated, and the results are exported. Otherwise, the procedure goes back to Step 3 for recurrence. The result is considered to be stable when the pressure drop $\Delta p$ across the system tends to extremum.
\end{enumerate}

\begin{figure}
  \centering
  \includegraphics[width=7.5cm]{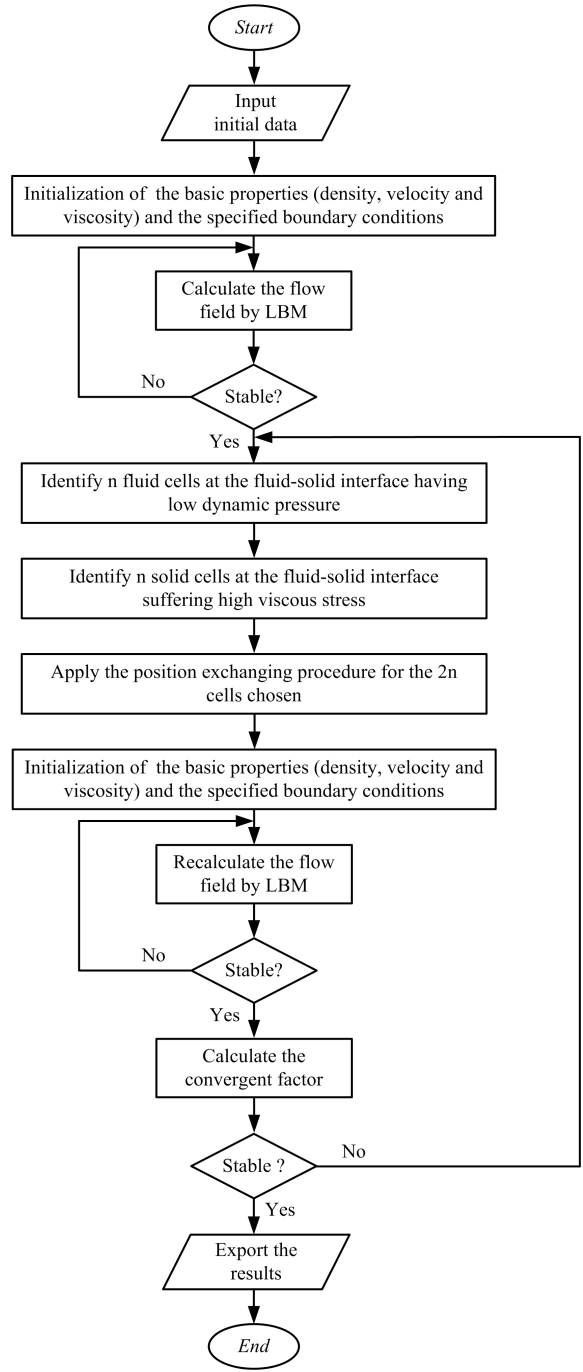}
  \caption{Flow chart of LBM based shape optimization algorithm}\label{fig2}
\end{figure}

We have tested an elemental T junction using our optimization algorithm, as shown in Fig. 3. It can be observed that T junction evolves into a Y shape junction with about 75\% decrease of $\Delta p$. It is expected that the total pressure drop of the arborescent structure having successive T junctions could also be further reduced by applying our algorithm.

\begin{figure}
  \centering
  \includegraphics[width=8.3cm]{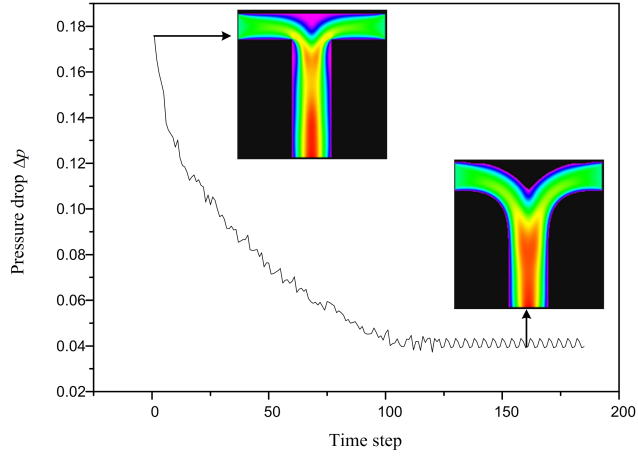}
  \caption{A simple numerical example of LBM based shape optimization algorithm}\label{fig3}
\end{figure}

\section{Design and optimization of flat type distributor}
\label{}
\subsection{Conventional flat type distributor}
Conventional flat type distributor comprises a simple cuboid distributor body having a fluid port at one face and a distribution surface at the opposite face, with the distribution surface having a plurality of uniformly spaced distribution openings. We choose the conventional flat type distributor as our comparison target.

\subsection{Arborescent distributor}
A flat type arborescent distributor is also designed for comparison. As shown in Fig. 4, the tree-shaped structure is induced by four generations of T-bifurcation or division, the number of outlets being $2^{4}=16$, uniformly distributed in a square surface. The four generations or scales are indexed from 0 to 4, with 0 for the single inlet channel and 4 for the smallest channels. Each outlet is 20 mm apart from its horizontal and vertical neighbors. For this reason, the channel length at each scale is employed as the ``local constraint", which obeys:
\begin{equation}\label{eq7}
l_{1}=l_{2}=20{\kern 2pt} \rm{mm}; {\kern 10pt}  \it{l}_{\rm{3}}=\it{l}_{\rm{4}}=\rm{10}{\kern 2pt} \rm{mm};
\end{equation}
\begin{figure}
  \centering
  \includegraphics[width=7cm]{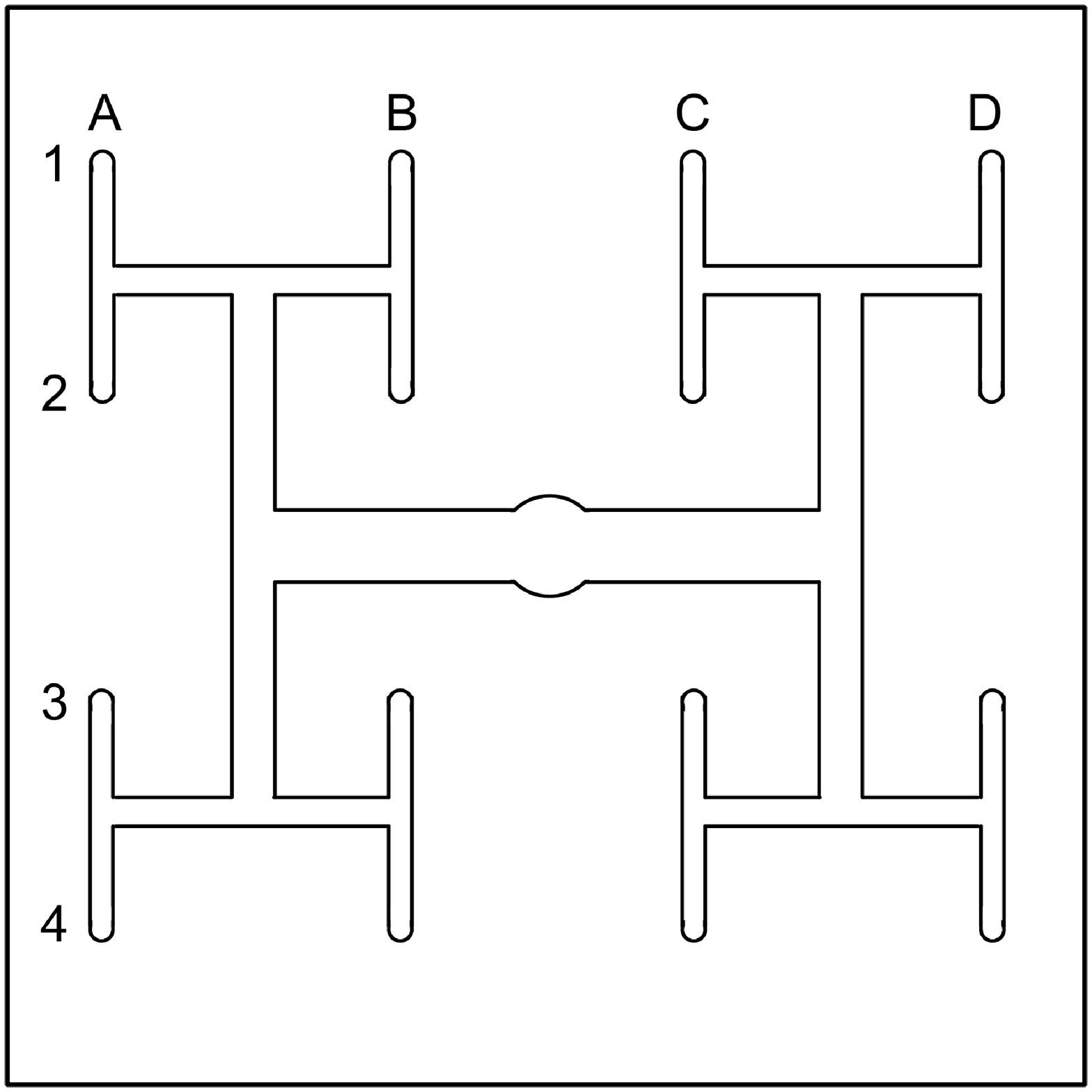}
  \caption{The arborescent structure of the flat-type distributor}\label{fig4}
\end{figure}
The cross-section of the distributor's channels is rectangle. The optimal scale ratio for the channel width $w_k$ at each scale is obtained by an optimization that accounts for both viscous dissipation and total pore volume, with the assumption of established laminar Poiseuille flow and neglecting the effects of flow singularities [26]. Detail optimization procedure may be found in the reference [27], and the scaling law for channel width:
\begin{equation}\label{eq8}
\left(\frac{{w_{k + 1} }}{{w_k }}\right)^4 \frac{{(h + w_k )(3h + w_k )}}{{(h + w_{k + 1} )(3h + w_{k + 1} )}} = \frac{1}{4}
\end{equation}
Where $h$  is the depth of the channels, which is considered as constant (10 mm) in a flat-type distributor.
In this work, $w_4$ is set to be 2 mm for the fabrication tolerance to be acceptable. With this specification, the dimensions for the bifurcated channels can be calculated using Eq. (9) and are listed in Table 1. Good flow distribution by this type of arborescent distributor has been reported because of its path symmetry that provides equivalent hydraulics characteristics, i.e. equivalent flow rate, equivalent time of passage and equivalent pressure drop [27,30].

\begin{table}
\centering
\caption{Dimensions of the arborescent distributor}
\label{table1}
\begin{center}
\begin{tabular}{@{}ccccc@{}} \toprule
\multicolumn{5}{c}{Channel} \\ \cmidrule(r){2-5}
Index & Number $n$ & Width $w$ & Depth $h$ & Length $l$  \\ \midrule
1 & 2 & 2.00 mm & 10 mm & 20 mm \\
2 & 4 & 2.91 mm & 10 mm & 20 mm \\
3 & 8 & 4.27 mm & 10 mm & 10 mm \\
4 & 16 & 6.35 mm & 10 mm  & 10 mm \\ \bottomrule
\end{tabular}
\end{center}
\end{table}

\subsection{Optimization by LBM based algorithm}
Based on the arborescent structure shown in Fig. 4, the entire computation domain is uniformly divided into elemental square sub-domains, which are categorized into fluid cells, solid cells, inlet cells and outlet cells (Fig. 5).
\begin{figure}
  \centering
  \includegraphics[width=9.3cm]{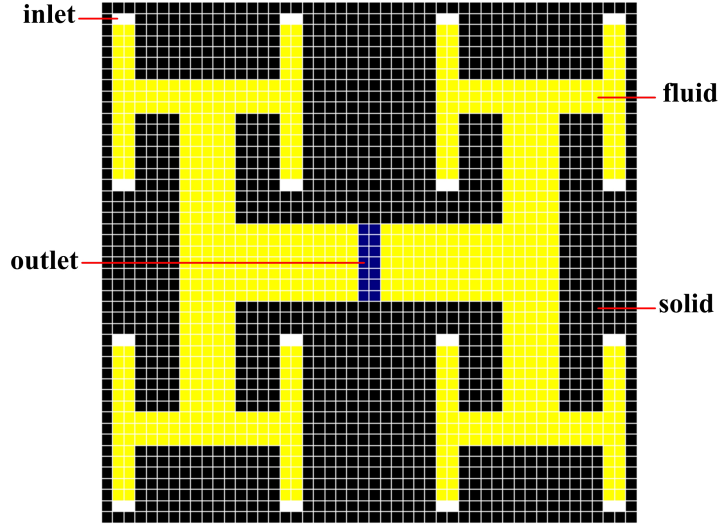}
  \caption{Cellular expression and boundary condition}\label{fig5}
\end{figure}
Considering the symmetry of simulation domain, half of the simulation domain is computed. The simulation domain is divided into $180\times360$ cells. The width of inlet and that of outlet are $W_{\rm{in}}$ = 11 cells and $W_{\rm{out}}$ = 52 cells, respectively. The parabolic velocity profiles are imposed at inlets where the flow rates are identical and the Neumann boundary condition [49] at the outlet is enforced. The wall boundary condition given in this paper is the bounce-back scheme, which is adaptable, robust and easy to be implemented for fluid flow in complicated geometries. The fluid density and kinematic viscosity are $\rho_{\rm{0}}$= 1.0 and $\nu$= 0.04, respectively. The initial condition is the equilibrium distribution of the inputs over the whole computational domain, using a constant density $\rho_{0}$= 1.0 and zero velocity. The cross-sectional average fluid velocities of inlets are $u_{\rm{in}}$ = 0.08, corresponding to a Reynolds number $Re_{\rm{in}} = \frac{W_{\rm{in}} \cdot u_{\rm{in}}}{\nu} = 11 0.08/0.04 = 22$.

The evolution of shape and flow field at different optimization time steps is shown in Fig. 6. With the algorithm proceeds, the flow singularity (right angle) gradually disappears and becomes smoother, implying the alleviation of singularity effect. Finally, it reaches a relative steady state, i.e. the shape of fluid flow stays almost unchanged with increasing time step. It can be observed that the streamline of velocity profile is well-kept at the final shape so that the flow turns slowly and continuously rather than abruptly.
\begin{figure*}
  \centering
  \includegraphics[width=12cm]{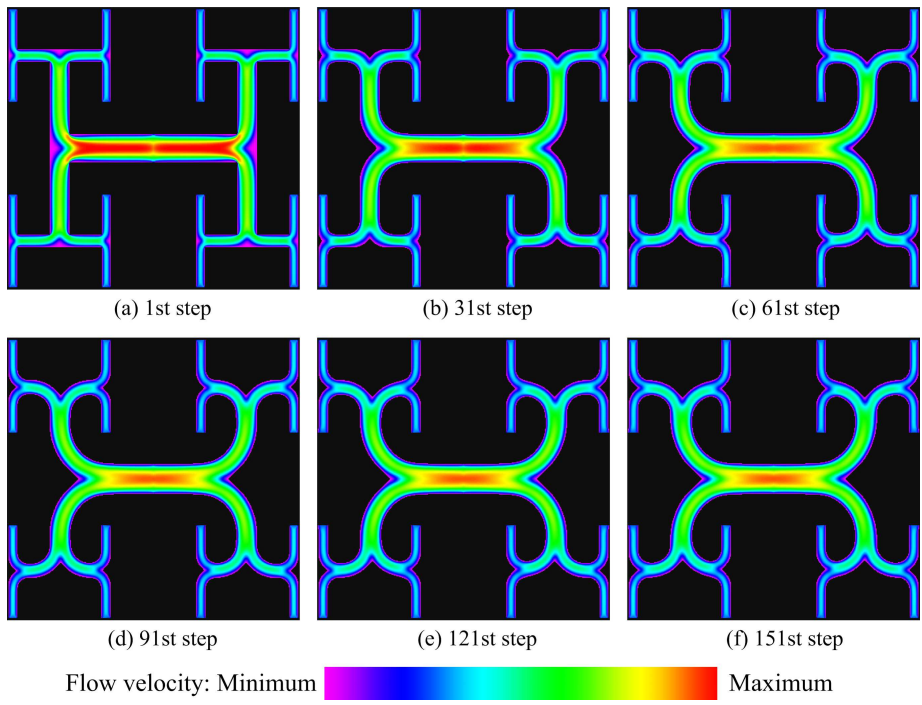}
  \caption{Shape and flow field evolution from first to the 151st step}\label{fig6}
\end{figure*}

It should be pointed out that what we optimized actually is an arborescent fluid collector for low Reynolds number applications. Standard LBM is developed for laminar flows at low Reynolds numbers, not applicable for problems with small kinematic viscosity as they associated with high Reynolds numbers, or being turbulent in many practical applications. Therefore, the incorporation of turbulence model into LBM together with heuristic optimality strategy to address practical systems at high Reynolds numbers will be well considered in our future work. Generally, there are two alternative ways to model turbulence: introducing the Reynolds equation and turbulent stress as in $k-\varepsilon$ model or using space-filtered governing equation and large eddy simulation (LES) with subgrid-scale stress model for the unresolved scale stress.

\subsection{Prototypes set-up}
Figure 7 shows the inner structures of conventional distributor, arborescent distributor and optimized distributor, respectively. The top has a thickness of 5 mm with a hole of 10 mm in the center connected to the inlet tube ($d_{\rm{in}}$=10 mm). The middle one having different structures . The bottom plate is 5 mm thick and has 16 holes ($d_{\rm{out}}$=2 mm) located at the 16 ends of the internal channels on the middle plate. Connecting to the holes are 16 tubes having an internal diameter of 2 mm and a length of 10 mm. They serve as the outlets of the distributor.
\begin{figure*}
  \centering
  \includegraphics[width=12cm]{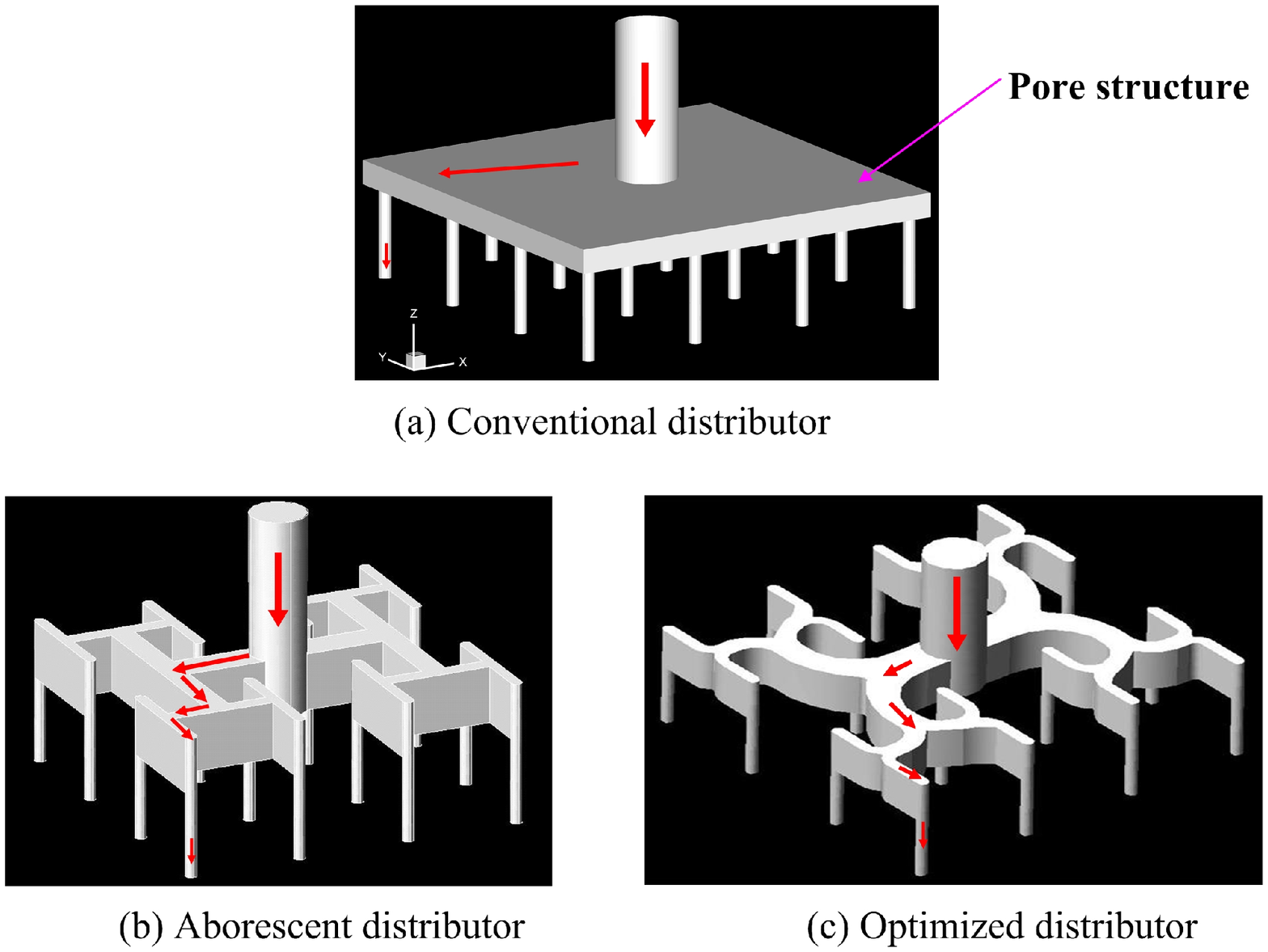}
  \caption{Pore structure of the flat-type distributors}\label{fig7}
\end{figure*}

\section{Numerical simulations}
The distributors together with the initial one, optimized by constructal laws and  by LBM optimality algorithm are assigned respectively as conventional distributor, arborescent distributor and optimized distributor, have been tested numerically, in order to investigate and compare their flow distribution and hydraulic performances.

\subsection{CFD simulation}
For the CFD simulation, the following assumptions were made:
\begin{enumerate}
  \item  [-] Fluid flow is incompressible, isothermal, Newtonian, in steady-state; the effect of viscous heating is negligible.
  \item  [-] The physical properties of the fluid (liquid water at 20$^\circ \mathrm{C}$) are: density $\rho=998.23$ $\rm{kg\cdot m^{-3}}$; viscosity $\mu=0.00101$ $\rm{kg\cdot m^{-1}\cdot s^{-1}}$. The operating pressure is set at 101325 Pa.
\end{enumerate}

The geometry model was built up using Hex-wedge elements by software GAMBIT 2.3.16. Note that with the symmetric assumption, one quarter of the real object was adopted as the model for the purposes of lessening the computational burden. Computational grids with $5.0\times10^{5}$ nodes were selectively refined in some local place where parametric variation was severe.

CFD commercial software FLUENT 6.3.16, of ANSYS, Inc., USA, was used here. Navier-Stokes equations were solved in 3D by standard $k-\varepsilon$ segregated turbulent solver with standard wall functions for near-wall treatment.The SIMPLE algorithm was employed for pressure-velocity coupling. The momentum equations were solved with second-order upwind differencing while the pressure terms were discretized using the standard scheme. At the inlet port, a constant velocity profile normal to the entry surface was used as initial boundary condition. The outlet ports were defined as pressure-outlet, with a zero relative pressure. A no-slip condition was imposed at the walls. The calculations were performed in double precision. The solution was considered to be converged when (i) the mass flow rate at each channel and the inlet static pressure were constant from one iteration to the next (less than 0.1\% variation) and (ii) the sums of the normalized residuals for control equations were all within the order of magnitude of $1.0\times10^{-6}$.

The simulations were conducted under the inlet water velocity ranging from 0.01 to 3.0 $\rm{m \cdot s^{-1}}$, corresponding to a Reynolds number within the range of 99.4$\sim$29821.0 in the inlet channel and an average Reynolds number within the range of 140.4$\sim$46679.0 in the outlet channel.

\subsection{Measurements of flow distribution uniformity}
Two dimensionless parameters, namely the maximal flow-rate ratio $\theta$ and the maldistribution factor $D_{\rm{g}}$, were used to evaluate the flow uniformity of distributors:
\begin{equation}\label{eq9}
\theta  = \frac{{\dot m_{\max } }}{{\dot m_{\min } }}
\end{equation}
\begin{equation}\label{eq10}
D_{\rm{g}}  = \sqrt {\frac{1}{{N - 1}}\sum\limits_{i = 1}^N {(\frac{{\dot m_i }}{{\dot m_{\rm{ave}} }}}  - 1)^2 }
\end{equation}
Where, $N$ stands for the sampling channels' number and $\dot m_{\rm{ave}}$ the mean flow rate of channels:
\begin{equation}\label{eq11}
\dot m_{\rm{ave}}  = \frac{1}{N}\sum\limits_{i = 1}^N {\dot m_i }
\end{equation}

\section{Discussion and Conclusion}
\subsection{Flow distribution}
CFD simulations have been performed to investigate the flow distribution performance of the initial and optimized distributors. Showing in Fig. 8 are the variation of maximal flow-rate ratio and the maldistribution factor for three distributors as a function of average Reynolds number at inlet. It can be observed that under the operating conditions examined, influence of flow-rate on the flow distribution uniformity for conventional distributor is large while this influence is very small for both arborescent distributor and optimized distributor. Both the maximal flow-rate ratio and maldistribution factor for conventional distributor are larger than those of both arborescent distributor and optimized distributor and decrease with the increasing average Reynolds numbers at inlet, implying that the fluid distribution performance of both arborescent distributor and optimized distributor are far better than that of conventional distributor. The flow maldistribution in both arborescent distributor and optimized distributor is within a very low level: the maximal flow-rate ratio less than 1.05 and the maldistribution factor less than 0.015. Note that the flow uniformity is a bit better for the arborescent distributor than the optimized one. The influence of the optimized algorithm on fluid distribution comes from two factors. On one hand, the optimized channel shape is not as symmetrical as before, at a bifurcation point, more fluid will go to the sub-branch with a small turning angle owing to the inertia of the fluid. And on the other hand the resolution of the square lattice for the LBM based optimized algorithm may be not high enough. The relative coarse boundary generated for the optimized construct would be transposed as input for geometrical model set-up for simulation, which may cause slight geometry differences and consequently very slight increase of flow maldistribution ($\Delta D_{\rm{g}} < 1\%$). Nevertheless, we emphasize that both distributors can guarantee excellent flow distribution uniformity under our operating conditions, with a large improvement with respect to the conventional case.

\begin{figure}
  \centering
  \includegraphics[width=8.3cm]{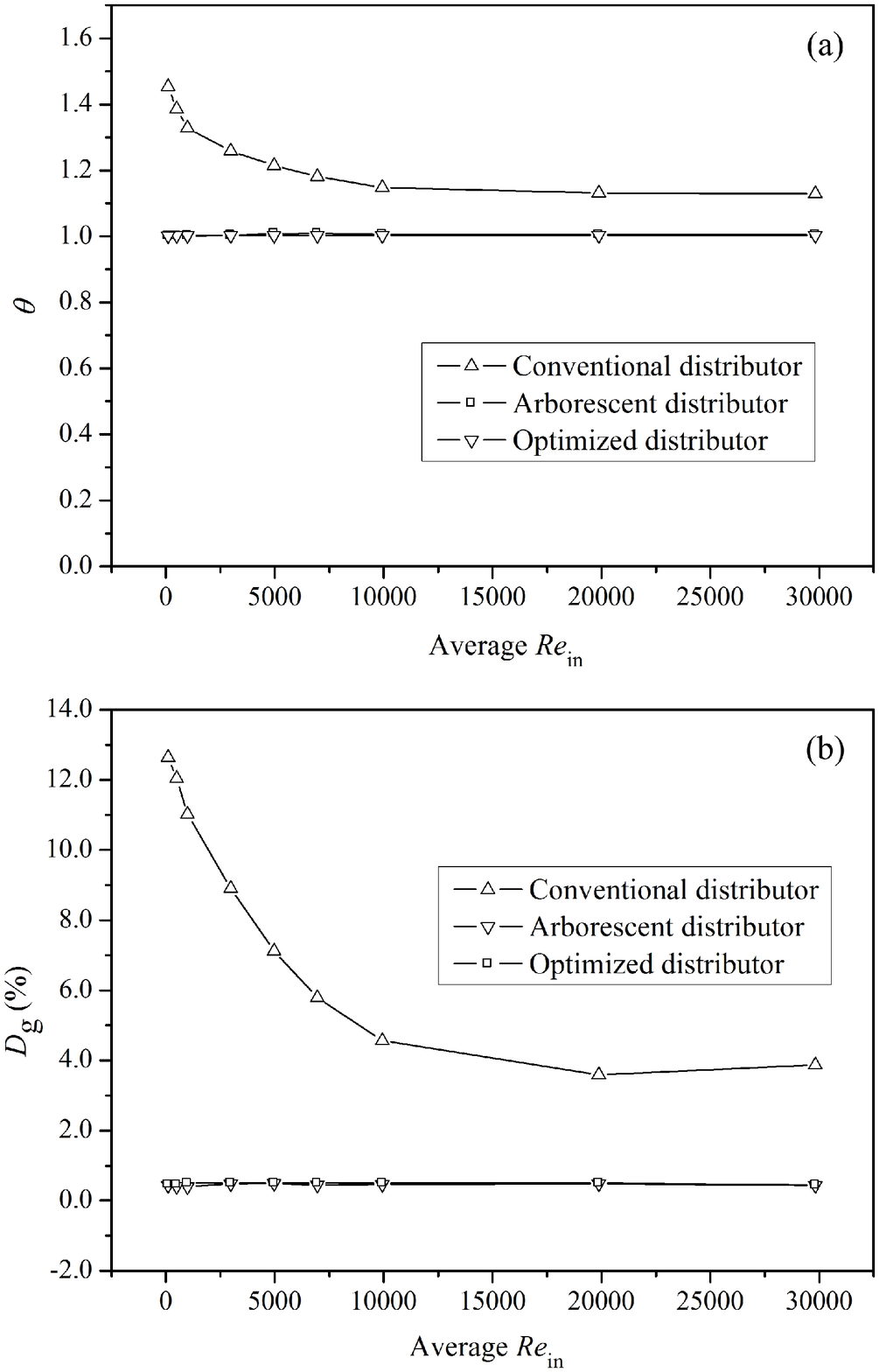}
  \caption{Flow distribution uniformity of the three distributors as a function of average Reynolds number at inlet. (a) maximal flow-rate ratio; (b) maldistribution factor}\label{fig8}
\end{figure}

\subsection{Pressure drop}
Figure 9 shows the contour of static pressure for the reference construct and for the optimized one. It can be clearly observed that there is a large pressure drop at the outlet of the distributor, which is more than 50\% of the total pressure drop. Besides the vertical inlet channel (scale 0), the pressure drops in the three distributors are generally located at the distributing network, and the frictional pressure drop in outlet parallel tubes is negligible. Examining the arborescent distributor, we observe that pressure drops are generally generated in T-type junctions where flow splits, because of strong viscous friction induced by the crash between the fluids and wall located in front of the flowing direction. This phenomenon has already observed and reported in our earlier work [34, 35]. Contrarily, the effects of singularities in the optimized case are largely eliminated because of the smooth transitions instead of the right angles in the Y-type junctions.
\begin{figure*}
  \centering
  \includegraphics[width=12cm]{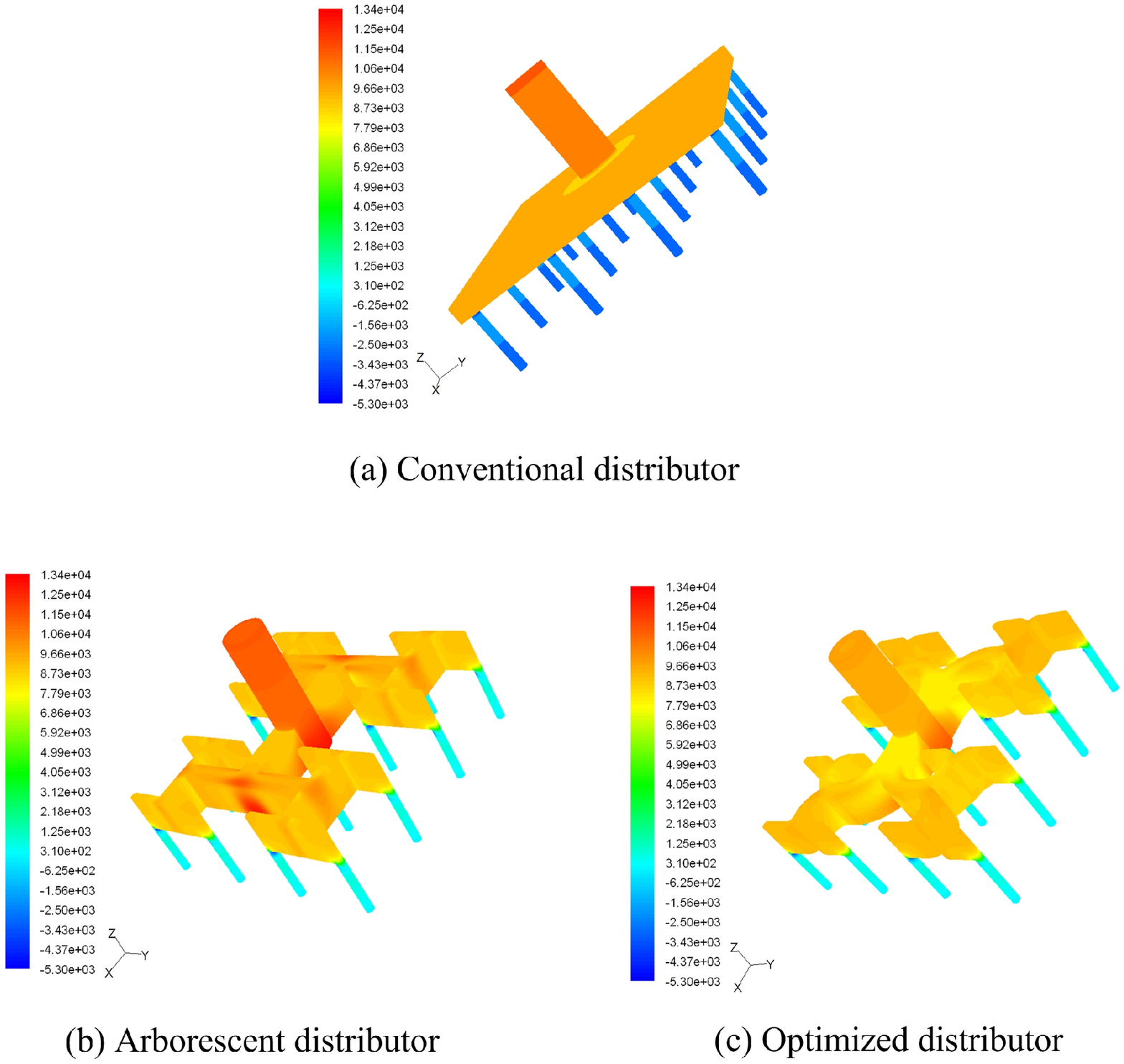}
  \caption{Contour of static pressure for three distributors (Inlet velocity: 2.0 $\rm{m\cdot s^{-1}}$; average $Re$ at inlet: 19881.0)}\label{fig9}
\end{figure*}

Figure 10 shows the comparison of pressure drop for the three distributors as a function of average Reynolds number at inlet. It is should be pointed out that the pressure drop of distributor is referred as the pressure difference between inlet tube connecting to the hole at top of distributor and 16 small outlet tubes connecting to the holes at bottom of distributor. The inlet static pressure of each distributor is within an acceptable range (less than 0.35 bar) under our tested conditions. The pressure drop of each distributor rapidly increases with increasing flow-rate. At the same flow-rate, the pressure drop of conventional distributor is far smaller than that of both arborescent distributor and optimized distributor. It should be noted that the pressure drop of arborescent distributor is consistently higher than that of optimized one because of its relatively higher flow resistance. The pressure drop in the distributor is reduced from 15.9\% to 25.1\% when average Reynolds number at the inlet increases from 99.4 to 29821.0, implying the good performance of our LBM based shape optimization algorithm in design and optimization of fluidic devises.
\begin{figure}
  \centering
  \includegraphics[width=8.3cm]{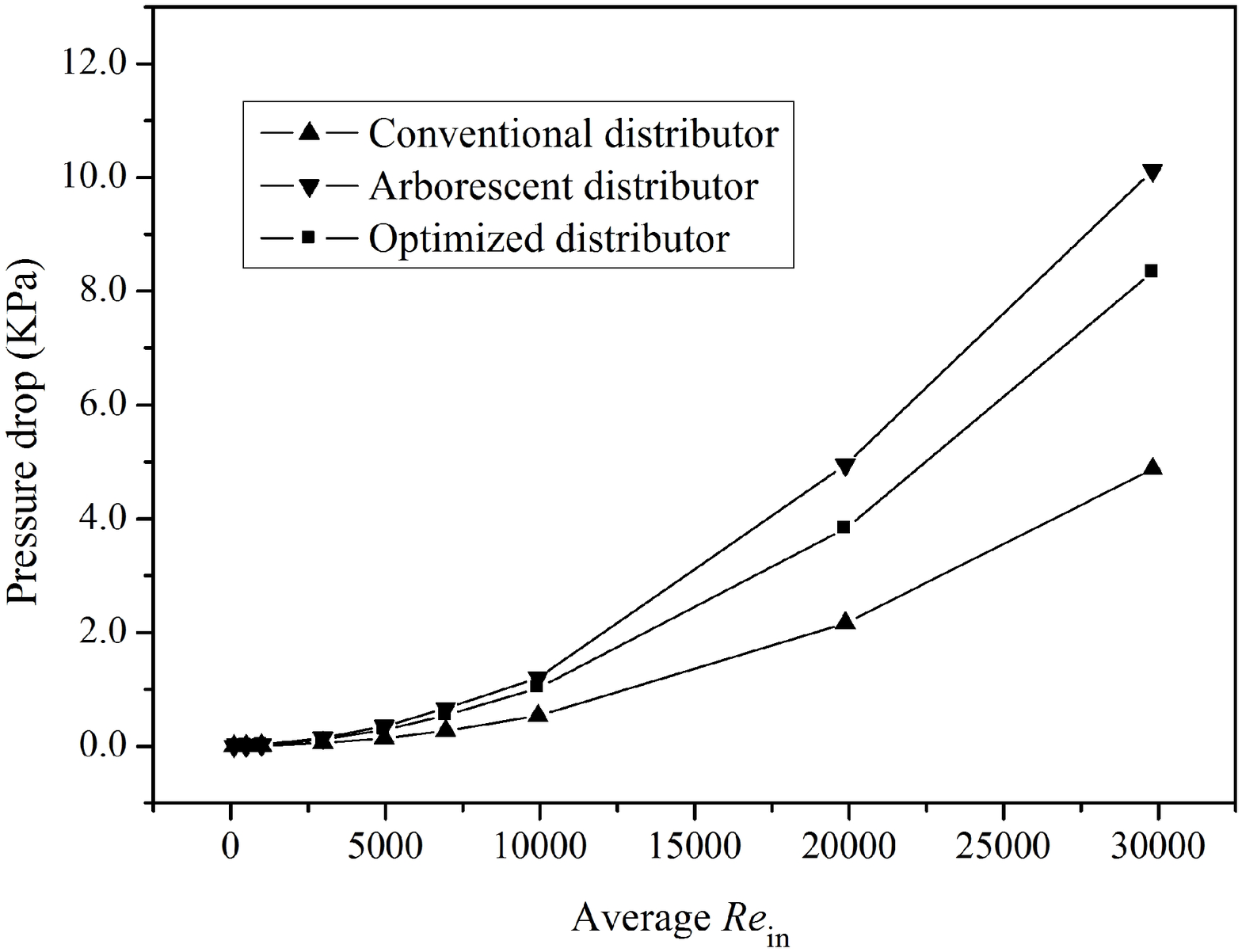}
  \caption{Pressure drop versus average $Re$ at inlet for three distributors (inlet velocity range: 0.01 - 3.0 $\rm{m\cdot s^{-1}}$)}\label{fig10}
\end{figure}

\section{Concluding remarks}
A flat-type arborescent distributor is optimized using a LBM based shape optimization algorithm developed by us. Distributor prototypes with conventional, arborescent and optimized arborescent structures are designed and investigated numerically. Numerical results show that both optimized distributor and arborescent distributor can guarantee excellence flow distribution uniformity while the conventional distributor increases maldistribution, the pressure drop of the conventional distributor is far lower than that of both arborescent distributor and optimized distributor and the pressure drop of the optimized flat-type distributor can be reduced between about 15.9\% and 25.1\% while keeping a uniform flow distribution, demonstrating the good performance of our optimization algorithm in design and optimization of complex structures of fluid flow.

Two different numerical methods, i.e. the discrete-velocity Boltzmann equation based CFD method (LBM) and the traditional Navier-Stokes based CFD method (Fluent) are used in this study. Indeed, our shape optimization algorithm is proposed as a method of searching for optimal architecture with minimal constraints. The limitation comes essentially from the initial morphology: there is no ``freedom" to ``morph" in the traditional Navier-Stokes based CFD as the geometry is fixed; there is a limited ``freedom" to ``morph" in classic analytical approaches (the establishment of scaling relations); and there is a relatively large ``freedom" to ``morph" with LBM approaches.

In general, the optimized distributor could find a balance between arborescent distributor and conventional distributor in both uniform distribution and pressure loss, and the pressure drop decrease is not as significant as expected with respect to the single T shape case (Fig. 3). Indeed, it would be more convenient by rounding the sharp corners in engineering cases. However, our algorithm actually tries to give an extreme of the pressure drop reduction. There might be two main reasons for this limited reduction. Firstly, this is due to the fact that the initial arborescent structure taken as the starting point for our optimization algorithm is already analytically optimized, as indicated by the established scaling relation of Eq. (9). What we searched for by applying the LBM based shape optimization algorithm is actually the ``further improvement" and the results obtained seem reasonable and encouraging. Secondly, the limiting factor lies in the 2D nature of our current optimization algorithm, i.e. the height of the channels is assumed infinite. In fact, the friction between the flow and the upper/bottom walls of channels with rectangular section, and the entrance effect (viscous dissipation in single inlet channel and in the first splitting T junction) contribute much to the total pressure drop in the flat-type distributor, as indicated by CFD simulation results (Fig. 9). A 3D optimization algorithm is expected by employing 3D lattice Boltzmann models, taking the influence of channel's cross-section shape (circular instead of rectangular) into account, for the optimization of 3D distributors presented in the references [26,29].

It should be pointed out that the optimized structure we obtained is a flat type arborescent fluid collector for low Reynolds number applications. However, CFD simulation results show that the performance improvement is obvious, even under turbulent flow conditions, indicating the robustness of the optimized architecture obtained using our algorithm. Incorporating proper turbulence model into LBM together with heuristic optimality strategy for high Reynolds number conditions will be another direction of our future work. Note that the decrease in total pressure drop might be even more evident for the use of collecting fluid flow (fluid collector), which is currently not able to be tested experimentally.

Finally, a further goal is to develop our algorithm for the optimization of fluid flow structures to optimize the performances of heat exchangers or reactors. In that case, it's clear that other heuristic optimality criteria that account for fluid flow, heat transfer or reaction kinetics should be introduced.

% \section{}
% \label{}
\nomenclature{$\textbf{c}$}{particle velocity, LT$^{-1}$}%
\nomenclature{$c_{\rm{s}}$}{speed of sound, LT$^{-1}$}%
\nomenclature{$\Delta p$}{pressure drop, L$^{-1}$MT$^{-2}$}%
\nomenclature{$d_{\rm{in}}$}{diameter of the single inlet channel, L}%
\nomenclature{$d_{\rm{out}}$}{diameter of the outlet tubes, L}%
\nomenclature{$h$}{channel depth, L}%
\nomenclature{$L$}{channel length, L}%
\nomenclature{$D_{\rm{g}}$}{maldistribution factor, dimensionless}%
\nomenclature{$\dot m_{\rm{ave}}$}{average flow-rate, MT$^{-1}$}%
\nomenclature{$\dot m_{\rm{max}}$}{maximum flow-rate between channels, MT$^{-1}$}%
\nomenclature{$\dot m_{\rm{min}}$}{minimum flow-rate between channels, MT$^{-1}$}%
\nomenclature{$n_{k}$}{number of channels in scale k, dimensionless}%
\nomenclature{$Re$}{Reynolds number, dimensionless}%
\nomenclature{$Re_{\rm{in}}$}{Reynolds number at inlet, dimensionless}%
\nomenclature{$\omega_i$}{lattice weighting factor, dimensionless}%
\nomenclature{$\textbf{u}$}{macroscopic velocity, LT$^{-1}$}%
\nomenclature{$\mu$}{fluid viscosity, L$^{-1}$MT$^{-1}$}%
\nomenclature{$\nu$}{kinematic viscosity, L$^{2}$T$^{-1}$}%
\nomenclature{$\rho$}{density, L$^{-3}$M}%
\nomenclature{$\rho_{\rm{0}}$}{mean density, L$^{-3}$M}%
\nomenclature{$\theta$}{maximal flow-rate ratio, dimensionless}%	
\nomenclature{$W_{\rm{in}}$}{width of inlet, L}%
\nomenclature{$W_{\rm{out}}$}{width of outlet, L}%
\nomenclature{$\tau_{yx}$}{shear stress in $x$ direction, L$^{-1}$MT$^{-2}$}%
\nomenclature{$\tau_{xy}$}{shear stress in $y$ direction, L$^{-1}$MT$^{-2}$}%
\nomenclature{$\Delta x$}{lattice spacing in $x$ direction, L}%
\nomenclature{$\Delta y$}{lattice spacing in $y$ direction, L}%
\nomenclature{$\Delta t$}{time interval, T}%
\nomenclature{$t$}{time step, T}%
\nomenclature{$f_i$}{particle distribution functions, dimensionless}%
\nomenclature{$f_i^{\rm{eq}}$}{particle equilibrium distribution function, dimensionless}%
\nomenclature{$\tau$}{relaxation time, T}%
\nomenclature{$\textbf{x}$}{particle position, L}%
\nomenclature{$u$}{velocity in $x$ direction, LT$^{-1}$}%
\nomenclature{$u_{\rm{in}}$}{velocity of inlet, LT$^{-1}$}%
\nomenclature{$\overline{u}_{\rm{out}}$}{cross-sectional average fluid velocity of outlet, LT$^{-1}$}%
\nomenclature{$\overline{u}$}{cross-sectional average fluid velocity, LT$^{-1}$}%
\nomenclature{$\upsilon$}{velocity in $y$ direction, LT$^{-1}$}%
\nomenclature{$p$}{pressure of fluid, L$^{-1}$MT$^{-2}$}%
\nomenclature{$q$}{dynamic pressure of fluid, L$^{-1}$MT$^{-2}$}%
\printnomenclature

\section*{Acronyms}
 \begin{acronym}[D3Q27]
 \begin{small}
 \acro{CA}{cellular automata}
 \acro{CFD}{computational fluid dynamics}
 \acro{D2Q9}{tow-dimensional nine-velocity lattice Boltzmann model}
 \acro{LBE}{lattice Boltzmann equation}
 \acro{LBM}{lattice Boltzmann method}
 \acro{LES}{large eddy simulation}
 \acro{LGA}{Lattice Gas Automata} 
 \end{small}
 \end{acronym}

%\acknowledgements % appendix
\section*{Acknowledgements}
This work is financially supported by the National Natural Science Foundation of China under Grants No. 21106155 and the Chinese Academy of Sciences under Grant No. XDA07080303. The authors would also like to thank Dr. Daniel Tondeur for fruitful discussions.



%% References


%\bibliographystyle{elsarticle-num}



\end{document}